\documentclass[oribibl]{llncs}
\usepackage{graphicx}
\usepackage{amsmath}
\usepackage{caption}
\usepackage{subcaption}
\usepackage{sidecap}
\usepackage{wrapfig,lipsum,booktabs}
\usepackage{chngcntr}
\usepackage[usenames]{color}

\makeatletter
\renewcommand*{\p@section}{\S\,}
\renewcommand*{\p@subsection}{\S\,}
\makeatother

\begin{document}

\title{An Empirical Study of Online Packet Scheduling Algorithms}
\author{Nourhan Sakr\thanks{n.sakr@columbia.edu} \and Cliff Stein\thanks{Supported in part by NSF grant CCF-1421161. cliff@ieor.columbia.edu}}

\institute{Industrial Engineering and Operations Research, Columbia University, NY 10027}
\maketitle

\begin{abstract}

This work studies online scheduling algorithms for
buffer management, develops new algorithms,
and analyzes their performances. Packets arrive at a release time
$r$, with a non-negative weight $w$ and an integer deadline $d$.
At each time step, at most one packet is
scheduled. The modified greedy (MG) algorithm is 1.618-competitive for
the objective of maximizing the sum of weights of packets sent, assuming
agreeable deadlines. We
analyze the empirical behavior of MG in a situation with arbitrary deadlines and demonstrate that it is at a
disadvantage when frequently preferring maximum weight packets over
early deadline ones. We develop the MLP algorithm, which remedies this
problem whilst mimicking the behavior of the offline algorithm. Our
comparative analysis shows that, although the competitive ratio of MLP is not as good as that of MG, it performs better in practice. We validate this by
simulating the behavior of both algorithms under a
spectrum of simulated parameter settings. Finally, we propose the design of
three additional algorithms, which may help in
improving performance in practice.

\end{abstract}

\section{Introduction}
\label{sec:Introduction}

Efficient buffer management at a network router is a critical issue
that motivates the online packet scheduling problem. Kesselman et
al.~\cite{kesselman04} introduce a buffer management delay model and
give algorithms to minimize end-to-end delay.  We adopt a similar
model to analyze the empirical behavior of the modified greedy (MG)
algorithm introduced in~\cite{jez12}, and propose new algorithms that do not have 
as strong worst-case guarantees, yet perform better in our
simulated settings.

\paragraph {Model.} For simplicity, we investigate a network router with
two nodes.   Studying a two node router is a first step towards understanding
more complicated and realistic models.
In~\ref{sec:MD} we briefly discuss possible model
modifications. At each integer time step, packets are buffered upon
arrival at the source node, then at most one packet is chosen from the
buffer to be sent to the target node. A packet ($r$,$d$,$w$) arrives
at a release date $r$, has a non-negative weight $w$, and needs to be
sent by an integer deadline $d$. A packet not sent by $d$ expires, and
is dropped from the buffer.  The objective of a packet-scheduling
algorithm $A$ is to maximize its weighted throughput, $\zeta_A$,
defined as the total weight of packets sent by $A$. It is easy to
relate our model to an online version of the classical offline  unit-job
scheduling problem where the input is a set of
$n$ unit-length jobs, each specified by a similar triple ($r$,$d$,$w$)
and the objective is to maximize weighted througput, that is the total weight 
of jobs that are processed before their deadlines.

\paragraph {Parameters.} 
We will typically be generating our input according to some type of
distribution. Let $T$ denote the number of time steps during which
the system can generate arriving packets, and let $\lambda$ denote an
arrival rate. We choose values for $T$ and $\lambda$. Then at 
each integer time step $t=1,\ldots,T$, we first
generate the number of arriving packets according to a Poisson
distribution with rate $\lambda$. For each arriving packet, we set
$r=t$ and generate $w$ from a uniform (integer) distribution
$U(1,w_{\max})$. To find $d$, we first generate
$\tau$, a time to expire, from a uniform (integer) distribution $U(0,d_{\max})$,
and set $d=r+\tau$. We call this Model 1. We also consider a bimodal distribution for $\tau$ with
weights $p$ and $1-p$, respectively, for two distinct distributions
centered on different means and call this Model 2. Although a network
may induce correlations between packets,  we use i.i.d. distributions as a first
step in modeling the behavior of our algorithms. 

In order to evaluate the performance of an online scheduling algorithm
(A), we use an offline algorithm (OFF) for comparison, which given all
future arrivals and packet characteristics, is able to statically find
the optimal solution (e.g. using maximum-weight bipartite
matching). Its solution gives the highest possible throughput the
system can achieve. The online algorithm is k-competitive if
$\zeta_{A}$ on any instance is at least 1/k of $\zeta_{OFF}$ on this
instance. The smallest k for which an algorithm is k-competitive is
called the competitive ratio~\cite{borodin98}. According
to~\cite{hajek01}, $k$ will be at most 2 for any algorithm that uses a
static priority policy. In this paper, we will simulate the online
algorithm and evaluate the ratio $\zeta_{A}$/$\zeta_{OFF}$. The
average of these ratios across each batch of simulations will be
denoted by $\rho_A$, where $A$ is the corresponding online algorithm.

\paragraph {Related Work.} The literature is rich with works that
acknowledge the importance of buffer management and present algorithms aiming at better router performance. Motivated by~\cite{kesselman04},~\cite{chin06}
gives a randomized algorithm, RMIX, while~\cite{bienkowski08} proves that it remains
$\frac{e}{e-1}$-competitive against an adaptive-online
adversary. Many researchers attempt to design algorithms
with improved competitive ratios. The best lower bound on the
competitive ratio of deterministic algorithms is the golden ratio
$\phi$~\cite{hajek01,chin03}. A simple greedy algorithm that schedules
a maximum-weight pending packet for an arbitrary deadline instance is 2-competitive~\cite{hajek01,kesselman04}. Chrobak et al.~\cite{chrobak04} introduce the first deterministic algorithm to have a competitive ratio strictly less than 2, namely 1.939. Li et al.~\cite{li07} use the idea of dummy packets in order to design the DP algorithm with competitive ratio at most 1.854. Independently,~\cite{englert07} gives a 1.828-competitive algorithm. Further research considers natural restrictions on packet deadlines with hopes of improving the competitive ratio. One type of restriction is the agreeable deadline model considered
in~\cite{jez12}, i.e. deadlines are (weakly) increasing in their
release times. Motivated by a more general
greedy algorithm, $EDF_\alpha$~\cite{chin06}, that schedules the earliest-deadline
pending packet with weight at least $1/\alpha$ of the maximum-weight
pending packet,~\cite{jez12}  develop the MG algorithm
which will be described in~\ref{sec:MG}. In other models, researchers enforce the FIFO discipline using a model where packets have no deadlines and the buffer is finite. One of the earliest such algorithms is the FIFO preemptive model studied by~\cite{andelman03}. Works such as~\cite{englert06,kesselman05,lotker02} adopt similar ideas. We do not consider the FIFO discipline in this paper.


\paragraph{Our Contribution.} We observe that while MG is $\phi$-competitive for the case of agreeable deadlines, it may not be the best option
to apply in practice. We demonstrate  the undesirable
performance of MG under certain scenarios, e.g. frequently preferring
maximum weight (late deadline) packets over early deadline ones. Our
proposed MLP algorithm remedies this drawback, as it outperforms MG on
most simulated instances. However, we are able to develop hard
instances to prove that MLP does not provide better worst-case
guarantees, whereas on those instances MG would produce the same
results as an offline solution. Contrasting the advantages of MG and
MLP motivates us to explore further algorithmic adjustments which may
improve performance, at least in practice, as supported by our
preliminary analysis. Finally, we justify that a two-node model with
an infinite buffer is a sufficient model for our
analysis. Moreover, extending the model to multiple nodes or imposing a
threshold on the capacity of the buffer does not significantly alter the performance
of the online algorithms.  

\section{Modified Greedy Algorithm (MG)}
\label{sec:MG}

MG is a $\phi$-competitive deterministic online algorithm for the
agreeable deadline model~\cite{jez12}. It focuses on two packets at
each time step: the earliest deadline non-dominated packet $e$
(i.e. maximum weight among all earliest-deadline packets) and the
maximum weight non-dominated packet $h$ (i.e. earliest deadline among
all maximum-weight packets in the buffer). Packet $e$ is chosen if
$w_e \geq \frac{w_h}{\phi}$ ($\phi\approx 1.618$) and $h$ is chosen
otherwise. While~\cite{jez12} consider an agreeable deadline model,
we relax this assumption and explore MG in a more general setting.

\paragraph{ MG analysis.} Although MG has the best competitive ratio
among deterministic online algorithms, we believe
that by better understanding MG, we can improve on it in
practice. Intuitively, if MG, at early stages, chooses packets with
longer deadlines (due to higher weights) over those with early
deadlines, then as time passes, many early packets expire while
most of the heavy later-deadline  packets will have already been
sent. Therefore, the algorithm may resort to choosing packets
with even smaller weights, thereby
wasting an opportunity to send a higher weight packet that has
already expired.  We, hence, explore the decisions made by MG by observing
its relative frequency of choosing $h$ over $e$.  

In order to consider a diverse set of instances, we set $T$ to 200 and
define ranges [0.7,20], [1,20] and [1,40] for $\lambda$, $w_{max}$ and
$d_{max}$, respectively. Under the assumptions of Model 1, we run a
batch of 200 simulations each for 8000 sampled parameter combinations. Given
each parameter combination, we calculate the relative frequency of
choosing $h$ over $e$ and average the frequencies over $\lambda$ to
obtain the empirical probability of $P(w_h>1.618 w_e)$, denoted by
$\psi$. 

\begin{wrapfigure}[15]{r}{0.45\textwidth}\centering
\vspace{-18pt}
\includegraphics[width=1.05\linewidth]{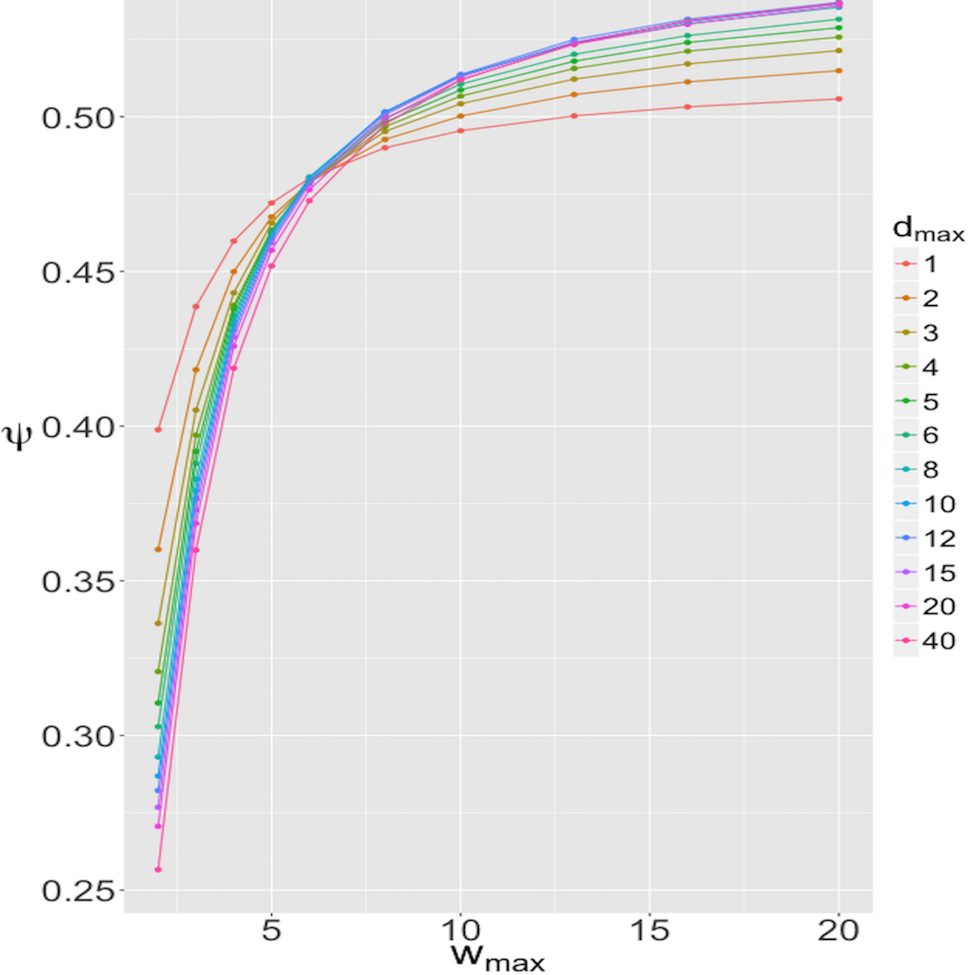} 
\caption{$\psi$ vs. $w_{max}$, colored by $d_{max}$}
\label{fig:MGdisadv}
\end{wrapfigure}

We suspect that when $\psi$ is high, especially if $h$ expires
at later deadlines, MG will  be at a major
disadvantage. Figure~\ref{fig:MGdisadv} plots $\psi$ vs. $w_{max}$,
where each curve corresponds to a fixed level for $d_{max}$. In
general, $\psi$ increases with $w_{max}$ and
$d_{max}$. The decreasing curve slope implies that $\psi$ is more
sensitive to lower values of $w_{max}$. Further analysis shows that at
any level of $w_{max}$, MG will
choose h over e at most 66\% of the time. We also observe that regardless the average number of
packets in the buffer, if $w_{max}$ is small (less than 3), 
the event of interest occurs at most 40\% of the time.


\begin{figure}[ht]\centering
\begin{subfigure}{.48\textwidth}
  \centering
  \includegraphics[width=\linewidth]{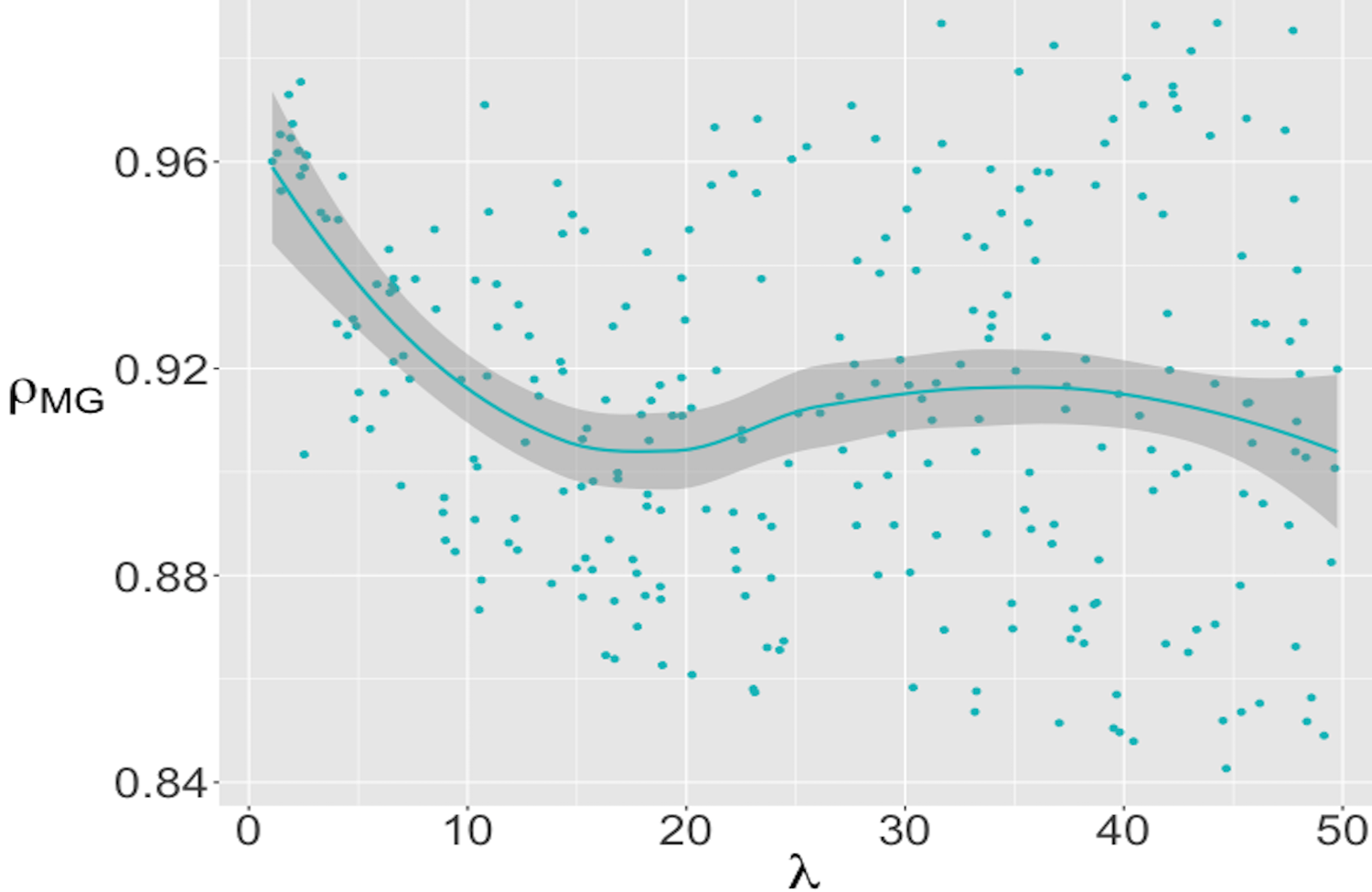}
  \caption{$\rho_{MG}$ vs. $\lambda$}
  \label{fig:MGtrapL}
\end{subfigure}
\begin{subfigure}{.48\textwidth}
  \centering
  \includegraphics[width=\linewidth]{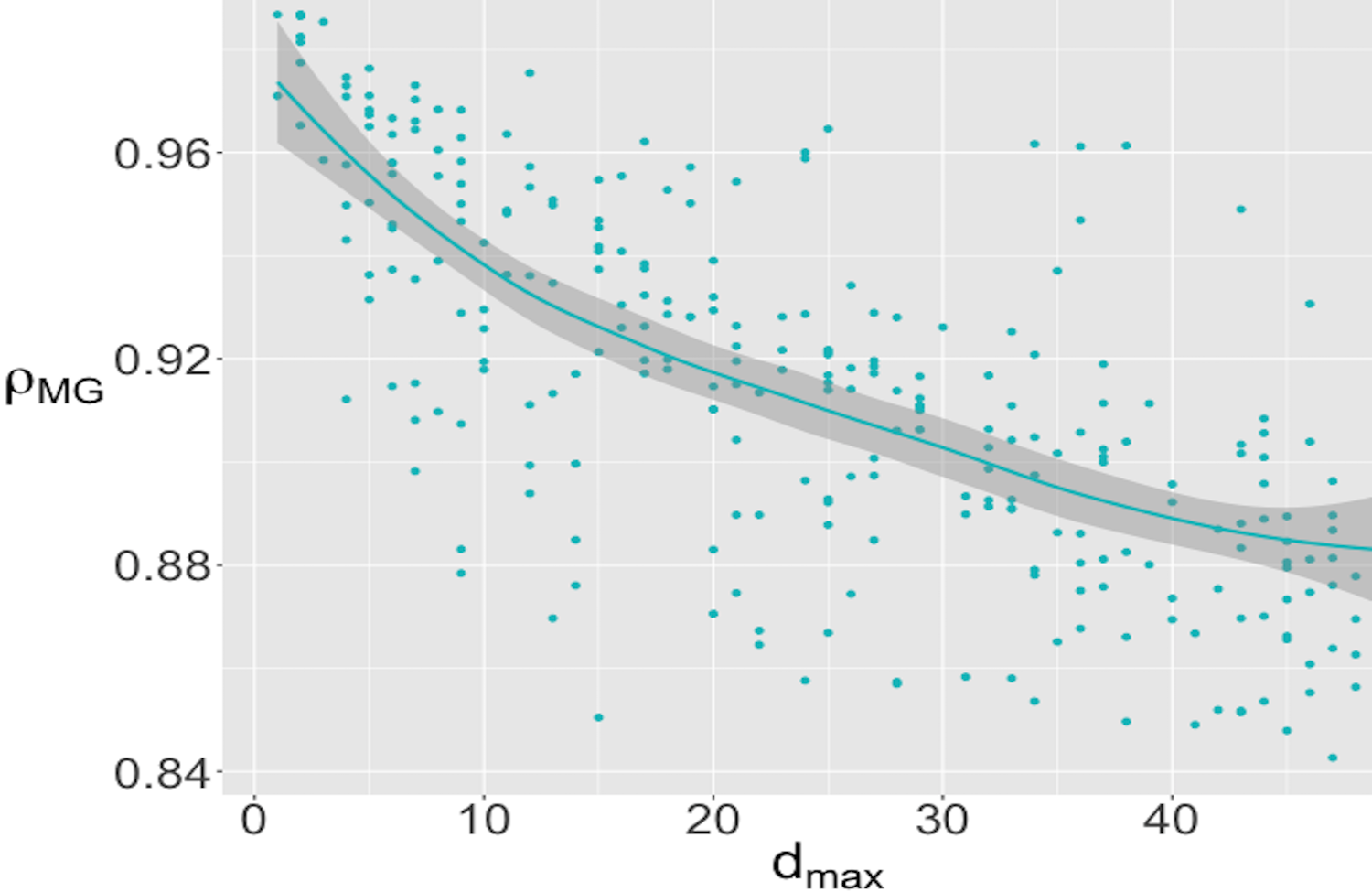}
  \caption{$\rho_{MG}$ vs. $d_{max}$}
  \label{fig:MGtrapD}
\end{subfigure}
\caption{Performance of MG under Scenario 1}
\label{fig:MGtrap}
\end{figure}

\vspace{-24pt} 
From this probability analysis, we conclude that unless
$w_{max}$ or $\lambda$ are small, MG tends to choose packet $h$ too
frequently. To show that this property may "fire back", we construct
Scenario 1, forcing MG to favor later deadlines. We reuse the data of
the generated packets above and adjust the weight of each packet by
multiplying it by its deadline. We let MG run on the new data and plot
$\rho_{MG}$ against different parameters. Figure~\ref{fig:MGtrap}
depicts lower ratios for the new dataset. While $\rho_{MG}$ originally
increased with $\lambda$, it now decreases with $\lambda$ and
$d_{max}$. $w_{max}$ does not affect the performance much. A gradient
boosted tree predictive model (i.e. a sequence of decision tree models
where the next model is built upon the residuals of the previous one)
shows that $d_{max}$ is the most important factor under Scenario 1, as
it accounts for 30\% of the variability in the model.

\section {Mini LP Algorithm (MLP)}
\label{sec:MLP}
\paragraph{Description.} In light of the previous analysis, we 
develop a new online algorithm that is more likely to send early
deadline packets.  The mini LP Algorithm (MLP) runs a "mini"
assignment LP at each time step in order to find the optimal schedule
for the current content of the buffer, assuming no more
arrivals. Assuming $n_t$ is the number of packets in the buffer at current time $t$, we search for the packet with the latest deadline ($d_{t,max}$) and set a timeline from $\hat{t}$= 0 to $d_{t,max}-t$. We then solve the following optimization problem, where $w_i$ is the weight of packet $i$ and $x_{i\hat{t}}$ is 1 if packet $i$ is sent at time $\hat{t}$ and 0 otherwise:
{\small \begin{eqnarray*}
\min  \sum_{i,\hat{t}} w_i x_{i\hat{t}}  & & \\
s.t. \;\;\; 
\sum_{i} x_{i\hat{t}} &\le 1 &  \;\;\; \hat{t}= 0,\ldots, d_{t,max}-t\\
 \sum_{\hat{t}} x_{i\hat{t}} & \le 1  & \;\;\; i=1,\ldots, n_t \\
 x_{i\hat{t}} &\ge 0 & \;\;\; i=1,\ldots, n_t
\end{eqnarray*}}
MLP then uses the optimal solution to send the packet that receives the first assignment, i.e. the packet $i$ for which $x_{i0}$=1, while
the rest of the schedule is ignored and recomputed in subsequent time
steps.

\subsection{Initial Analysis}
\label{subsec:MLP}

Similar to the MG analysis, we compute $\rho_{MLP}$ and are interested
in its behavior as the load varies. A way to measure load is to define the average number of packets in the buffer as
$\bar{n}$, which is a byproduct of $\lambda$. We expect higher
$\rho_{MLP}$ at low $\bar{n}$, since the online algorithm would not have
many packets to choose from and hence, is more likely to
choose the same packet as the offline algorithm at each time
step. However, we expect $\rho_{MLP}$ to decrease as $\bar{n}$ increases, since the discrepancy between online
and offline solutions increases. To test this, we sample parameter combinations from $T\in [100,500]$, $\lambda \in [0.7,20]$ and $w_{max}$, $d_{max}\in[1,20]$ and run a batch of 1000 simulations per
combination. 

\begin{wrapfigure}[13]{r}{0.48\linewidth}
\vspace{-10pt}
\includegraphics[width=1.15\linewidth]{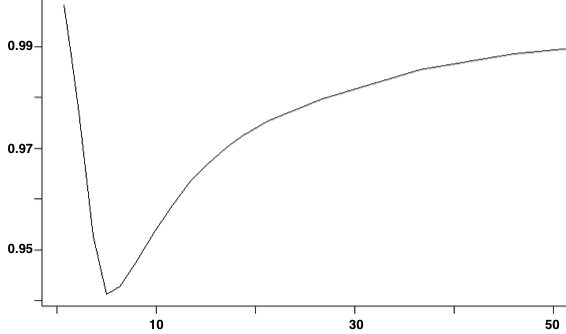}
\caption{$\rho_{MLP}$ vs. $\bar{n}$}
\label{fig:MLP1}
\end{wrapfigure}

Figure~\ref{fig:MLP1} plots $\rho_{MLP}$ vs. $\bar{n}$
and interestingly shows a dip-shaped graph: $\rho_{MLP}$ starts at a very high value ($\approx1$),
decreases as expected with increasing $\bar{n}$ until
eventually it increases again, thereby forming a dip, and finally
converges to 1. Our claim is true at first,
when $\lambda$ is relatively low, as the first range for $\lambda$ is
quite sensitive ($\lambda=2.3$ vs. 2.8 makes a difference). However,
when $\lambda$ increases, the problem loses its sensitivity. An
explanation for such behavior may be that as $\bar{n}$ increases (with
increasing $\lambda$), we are more likely to have multiple packets
achieving maximum weight, in which case both the online and offline
algorithms are likely to choose those packets and have less
discrepancy between their choices, especially if the weight or
deadline ranges are not wide enough. We conclude that when the system
is heavily or lightly loaded, both algorithms perform well. The dip
happens in the interesting area. Consequently, we will investigate 
how the dip moves and what the effect of parameter choices will have
on such graph.

\subsection {Parameter Effect on MLP Behavior}
Changing  the parameters to generate different graphs
did not change the structure of the dip-shaped graph that we have seen in Figure~\ref{fig:MLP1}. Nonetheless, the
dip gets narrower/wider and shifts to the left/right, as parameters
change.  In this
section, we will only focus on a restricted range for the values of
$\lambda$, namely 0.7 to 10. However, we believe that the restriction does not
mask any interesting results, since MLP converges at higher values
of $\lambda$, as we have seen before. Therefore, a heavily loaded
system is not significant for our analysis.   

\paragraph{Arrival Rates.} The graph inevitably depends on $\lambda$,
as it directly affects $\bar{n}$, i.e. the x-axis. However,
$\lambda$ does not have a direct effect on the shape of the graph. By tuning the range for $\lambda$, we are able to ``zoom in''
onto the dip area and monitor the behavior more accurately where the
system is neither lightly nor heavily loaded. The range for such
sensitive values is on average between 1.3 and
4.2. Figure~\ref{fig:MLP2} (in the appendix) zooms in on the dip where $\lambda$ is most
sensitive.

\paragraph{Weight Ranges.} The range of the weights moves the dip to
the right (left), as it gets narrower (wider). Very narrow ranges
(i.e. low values for $w_{max}$) are the most influential. As $w_{max}$
increases, its impact decreases. In fact, this result seems intuitive
and one can see an example in Figure~\ref{fig:MLP3} where the weight
range is designed to be very narrow ($w_{max}$ is set at
2). Some
experimentation led us to the explanation of this phenomenon: When
there are few options for weights, both algorithms converge
together. Let's say weights are only 1 and 2, then the higher the
$\bar{n}$, the more likely we will have packets of weight 2. In this
case both algorithms find the optimal choice to be the packet with
higher weight (we don't have much choice here so it must be 2). Hence,
both behave alike. We note that it is not in particular the range of
weights that has this effect but rather the number of distinct weights
available, i.e. choosing between weights 1 and 2 vs. 100 and 200,
would depict the same behavior.

\paragraph{Time Period and Deadline Range.} $T$ and $d_{max}$ have a
combined effect. Figures~\ref{fig:MLP4} and~\ref{fig:MLP5} give two
examples: Allowing a longer timeline $T$ results in a second but
higher dip and slows down convergence, such that suddenly higher
values of $\lambda$ become slightly more interesting. Meanwhile lower
$d_{max}$ values (combined with shorter $T$'s) result in a graph with
one sharp dip as well as much faster convergence to 1.

\subsection{Influence of maximum-weight packets}
\begin{figure}
\begin{subfigure}{.47\textwidth}
  \centering
\includegraphics[width=1.1\linewidth]{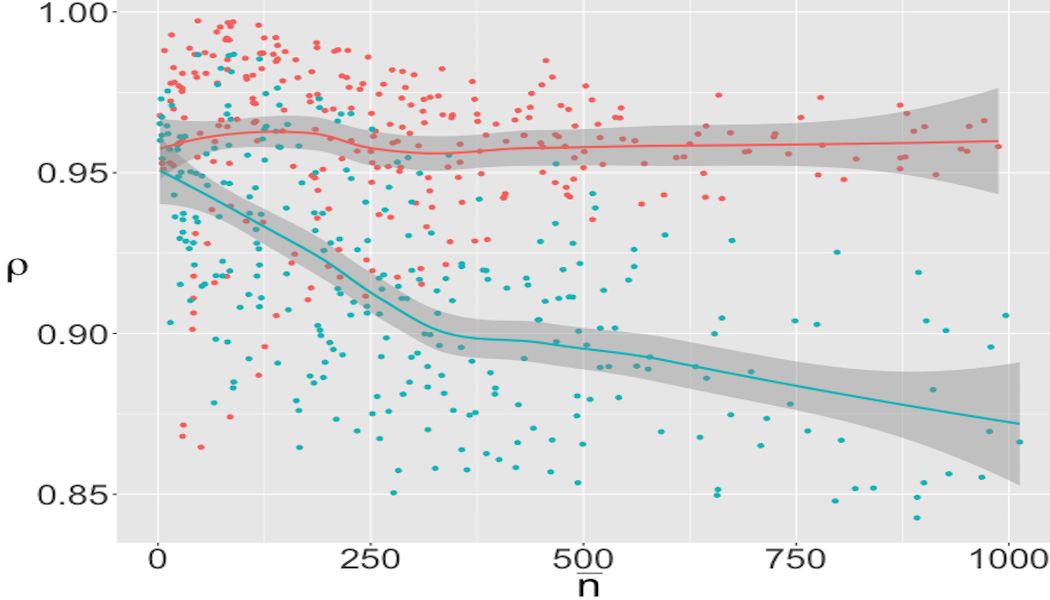}
\caption{$\rho$ vs. $\bar{n}$ under Scenario 1}
\label{fig:Scenario3}
\end{subfigure}
\hfill
\begin{subfigure}{.47\textwidth}
  \centering 
\includegraphics[width=1.1\linewidth]{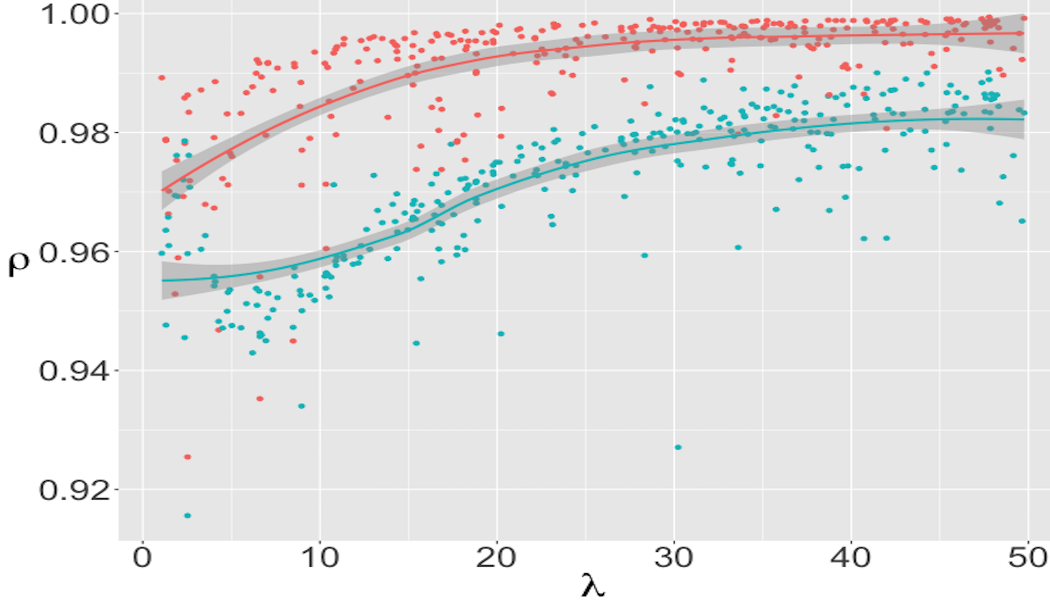}
\caption{$\rho$ vs. $\lambda$ under Scenario 2}
\label{fig:CompL}
\end{subfigure}
\caption{$\rho_{MLP}$(red) and $\rho_{MG}$(green)} 
\end{figure}

The motivation of MLP was mainly to remedy the drawback we observed
for MG when later deadline packets are preferred. Therefore, it is
essential to verify that MLP outperforms MG under Scenario 1. In fact,
one-sided 99\% confidence intervals (CI) imply that $\rho_{MG}$ is at
most 91.98\% while $\rho_{MLP}$ is at most 96.28\%. The difference in
performance between both algorithms increases with
$\lambda$. Figure~\ref{fig:Scenario3} shows the behavior of $\rho$
against $\bar{n}$ for both algorithms. While MLP is not influenced by
$\bar{n}$ under this scenario, the performance of MG gets worse as
$\bar{n}$ increases. A 99\% two-sided CI for
$\frac{\rho_{MLP}}{\rho_{MG}}$, denoted by $\hat{\rho}$, is
(1.0443,1.0552), implying that MLP produces a total weight at least
4.43\% more than that of MG under this scenario. Better performance is
observed with higher $T$ or lower $d_{max}$, but
$w_{max}$ does not seem to influence the algorithms' performances.
\vspace{-13pt}

\vspace{-30pt}
\section{Comparative Analysis}
\label{sec:CA}
\vspace{-8pt}
In this section, we contrast the behavior of MG and MLP under a
spectrum of parameter settings. We are interested in the behavior of
the ratios against our parameters and expect MLP to perform better in
our simulations. The general procedure for our simulations is based on
sampling parameter combinations from a predefined parameter space. 
we impose the following parameter range restrictions: $T \in
(50,750)$, $\lambda \in (0.5,50)$, $w_{max} \in (2,50)$ and $d_{max}
\in (1,50)$ (Scenario 2). For
each combination, we run MG, MLP (5 times each) and the offline
algorithm in order to obtain values for $\rho_{MG}$, $\rho_{MLP}$, as
well as $\hat{\rho}=\frac{\zeta_{MLP}}{ \zeta_{MG}}$. Detailed steps
for simulations are given in~\ref{app:CA}. 

\subsection {Ratio behavior w.r.t. model parameters}
Comparing $\rho_{MLP}$ and $\rho_{MG}$ against values of $\lambda$ implies that on average MLP outperforms MG (Figure~\ref{fig:CompL}). As $\lambda$ increases, both algorithms perform better. A 99\% one-sided CI for $\rho_{MG}$ is (0,0.9734), implying that we are 99\% confident that $\zeta_{MG}$ is at most 97.34\% of $\zeta_{OFF}$, while the one-sided CI for $\rho_{MLP}$ is (0,0.9926). In Figure~\ref{fig:COMP}, it is evident that MG produces a wider spread of the ratios. All else constant, the performance of each algorithm improves with higher $T$, lower $d_{max}$ or higher $\bar{n}$, whereas it is not influenced by the values of $w_{max}$. 

\begin{figure}[ht]\centering 
\vspace{-13pt}
\begin{subfigure}{.48\textwidth}
  \centering
\includegraphics[width=1.28\linewidth]{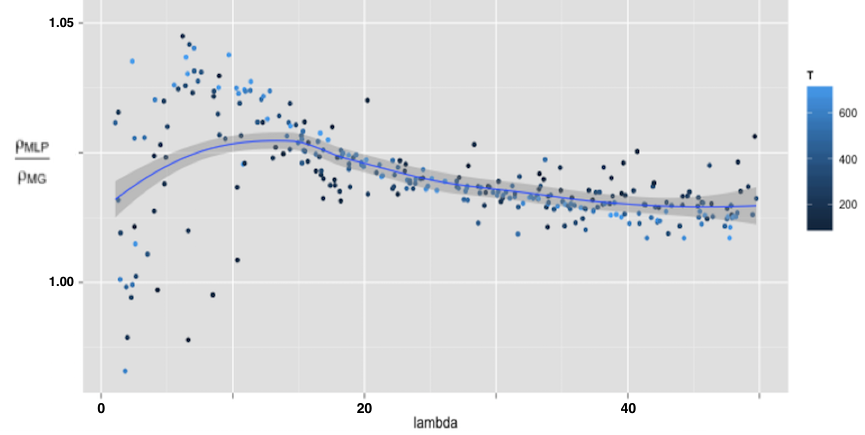}
\caption{under Scenario 2}
\label{fig:MLPtoMG}
\end{subfigure}
\hfill
\begin{subfigure}{.48\textwidth}
  \centering
\includegraphics[width=1.28\linewidth]{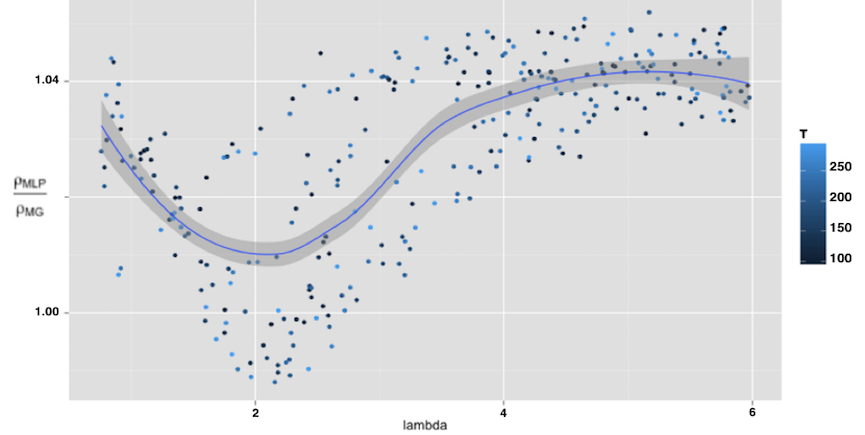}
\caption{under Scenario 3}
\label{fig:MLPtoMGtau}
\end{subfigure}
\caption{$\hat{\rho}$ vs. $\lambda$ and colored by T}
\end{figure}


 Figure~\ref{fig:MLPtoMG} plots $\hat{\rho}$ vs. $\lambda$, colored by $T$. For very small $\lambda$'s, there is the
possibility that MLP and MG perform similarly; in some cases, MG
outperforms MLP, regardless of the value of $T$. However, for large
$\lambda$, MLP tends to outperform MG. A 99\% two-sided CI for
$\hat{\rho}$ is $(1.0188, 1.0216)$, implying that we are 99\%
confident that $\zeta_{MLP}$ is at least 1.88\% more
than $\zeta_{MG}$. However, both algorithms have similar performance as the
upper bound of the CI shows that $\zeta_{MLP}$ is at most
2.16\% more. Whether this is beneficial depends on the use case as
well as time constraints (see~\ref{sec:HI}).~\ref{app:predM} presents a brief analysis where we construct
gradient booted tree predictive models on the ratios for inference
purposes.

\vspace{-14pt}
\subsection{Changing the distribution of $\tau$}

So far, we have only considered uniform distributions, however, real
inputs are more complicated.  Here we make one step towards
modeling more realistic inputs and consider a 
$\tau$ that follows a bimodal distribution of two distinct peaks (recall Model 2); with probability
$p$, $\tau$ is $N(2,0.5^2)$ and with probability $1-p$, $\tau$ is
$N(8, 0.75^2)$. We restrict our parameters to the following ranges: $T
\in (100,300)$, 

\begin{figure}[hb]\centering 
\vspace{-20pt}
\includegraphics[width=0.75\linewidth]{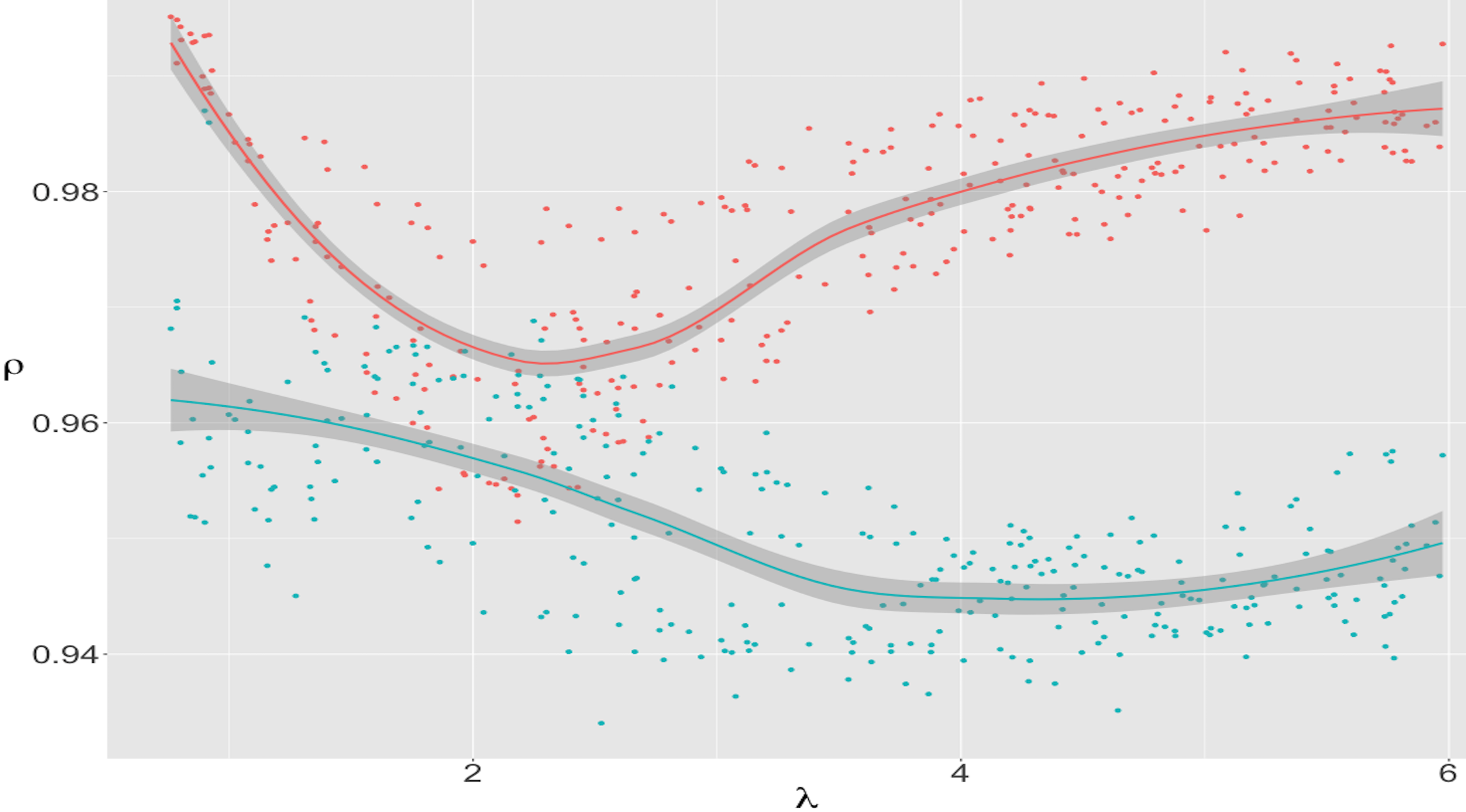}
\caption{$\rho_{MLP}$(red) and $\rho_{MG}$(green) vs. $\lambda$ under Scenario 3}
\label{fig:ChangeTau}
\end{figure}
 \noindent
$\lambda \in (0.7,6)$, $w_{max} \in (2,7)$ and $p \in
(0.75,0.95)$ (Scenario 3). 
We choose a bimodal distribution because these distributions are often
hard for scheduling algorithms.  Indeed, we see that the
results for Scenario 3 are slightly
different.

While MG performs worse with increasing $\lambda$, $\rho_{MLP}$
improves with $\lambda$ and still outperforms $\rho_{MG}$
(Figure~\ref{fig:ChangeTau}).The graph for $\rho_{MLP}$ resembles a
dip-shaped graph, yet we find this dip to be entirely above the
confidence interval of $\rho_{MG}$. All else constant, neither
algorithm is influenced greatly by any of the parameters $T$,
$d_{max}$ or $p$.  Figure~\ref{fig:MLPtoMGtau} plots $\hat{\rho}$
vs. $\lambda$, where lighter points correspond to longer $T$'s. For
very small $\lambda$, MLP and MG perform similarly. In some cases, MG
outperforms MLP, regardless of the value of $T$. However, for large
$\lambda$, a 95\% CI shows that MLP outperforms MG by at least 2.80\%
and at most 3.30\%.

\section{Hard Instances}
\label{sec:HI}
\begin{wrapfigure}[13]{r}{0.47\linewidth}
\vspace{-40pt}
\includegraphics[width=\linewidth]{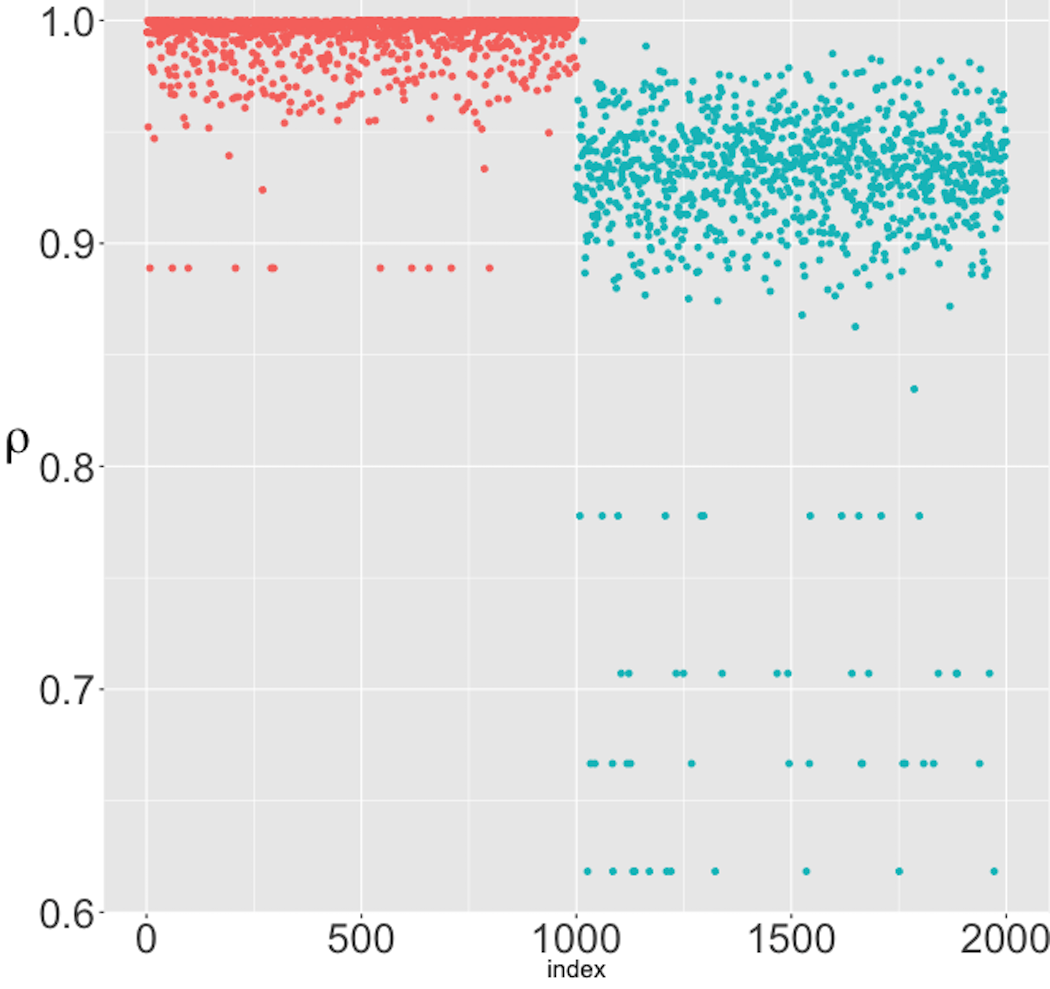} 
\caption{Index plot of $\rho_{MLP}$(red) and $\rho_{MG}$(green)}
\label{fig:Index}
\end{wrapfigure}

The previous analysis presents evidence  that MLP gives better
competitive ratios than MG.
An index plot
(Figure~\ref{fig:Index}) of $\rho_{MG}$ and $\rho_{MLP}$ shows that,
for the same instances, MLP not only outperforms MG, but also gives a
ratio of 1 for most of the instances that are hard for MG.  
However, it would be incorrect to conclude that MLP always has a
better competitive ratio than MG.  In fact, 
we are able to
create hard instances for MLP where it performs worse than MG. A small 
example is given in Table~\ref{HI-example}.

\vspace{5pt}
\begin{table}[ht] \centering
 \begin{tabular}{|r||r|r|r|} 
 \hline
 Packet(r,d,w) \;\;\; & MLP \;\; & MG \;\; & Offline \;\;   \\ [0.5ex] 
 \hline\hline
 $\mathbf{(1,1,w_1)}$ &   \;\;\; Assign to $t=1$  & & \\ 
 \hline
 $\mathbf{(1,2,w_2)}$ & & \;\;\; Assign to $t=1$ &  \;\;\; Assign to $t=1$ \\
 \hline
 $\mathbf{(2,2,w_2)}$ &  \;\;\; Assign to $t=2$ &  \;\;\; Assign to $t=2$ &  \;\;\; Assign to $t=2$\\  
 \hline
 \hline
 \;\; $\bf{Throughput}$  & $w_1+w_2$& 2$w_2$ & 2$w_2$\\ [1ex] 
 \hline
\end{tabular}
 \caption{Hard Instance for MLP}
 \label{HI-example}
\end{table}

\vspace{-16pt}
 For $w_2 > w_1$, we can easily see that MLP is 2-competitive, while
 $\rho_{MG}$ on those instances is 1. However, such worst-case
 instances may be rare and our results show that MLP performs better
 on a varied set of data.  On the other hand, MG, which bases its decisions on
 simple arithmetic, is simpler than MLP which solves an LP at each time
 step. In our experiments, MLP was as much as 140 times slower than MG.
In the next section, we consider some modifications to take advantage
of the strengths of both approaches.

\section{Algorithm Modifications}
\label{sec:AM}

 In an attempt to find faster solutions, we introduce some possible algorithmic
 modifications. Preliminary results show slight performance improvement when these
 modifications are applied. However, more analysis is needed to verify
 the results and choose the best parameters for improvement.

\subsection {The Mix and Match Algorithm (MM)}
\label{subsec:MM}

\paragraph{Algorithm.} The Mix and Match Algorithm (MM) combines 
both MG and MLP. At each time step, MM chooses to either
run MG or MLP-according to $\bar{n}$. If $\bar{n}$ is high, then by
previous analysis, MG and MLP each converges to 1 (assuming Model
1), and MM runs MG, as it is faster and has a competetive ratio that
is as good as that of MLP. If $\bar{n}$
is low, MM runs MLP, as it is more accurate and the running time is
also small since $\bar{n}$ is low.
To distinguish between "high" and "low", we define a
threshold $\bar{N}$. Although MM suffers from the same limitations as
MLP, it might still be preferred due to its smaller computation time. 

\begin{figure}[h!]\centering
\begin{subfigure}{.46\textwidth}
  \centering
  \includegraphics[width=1.2\linewidth]{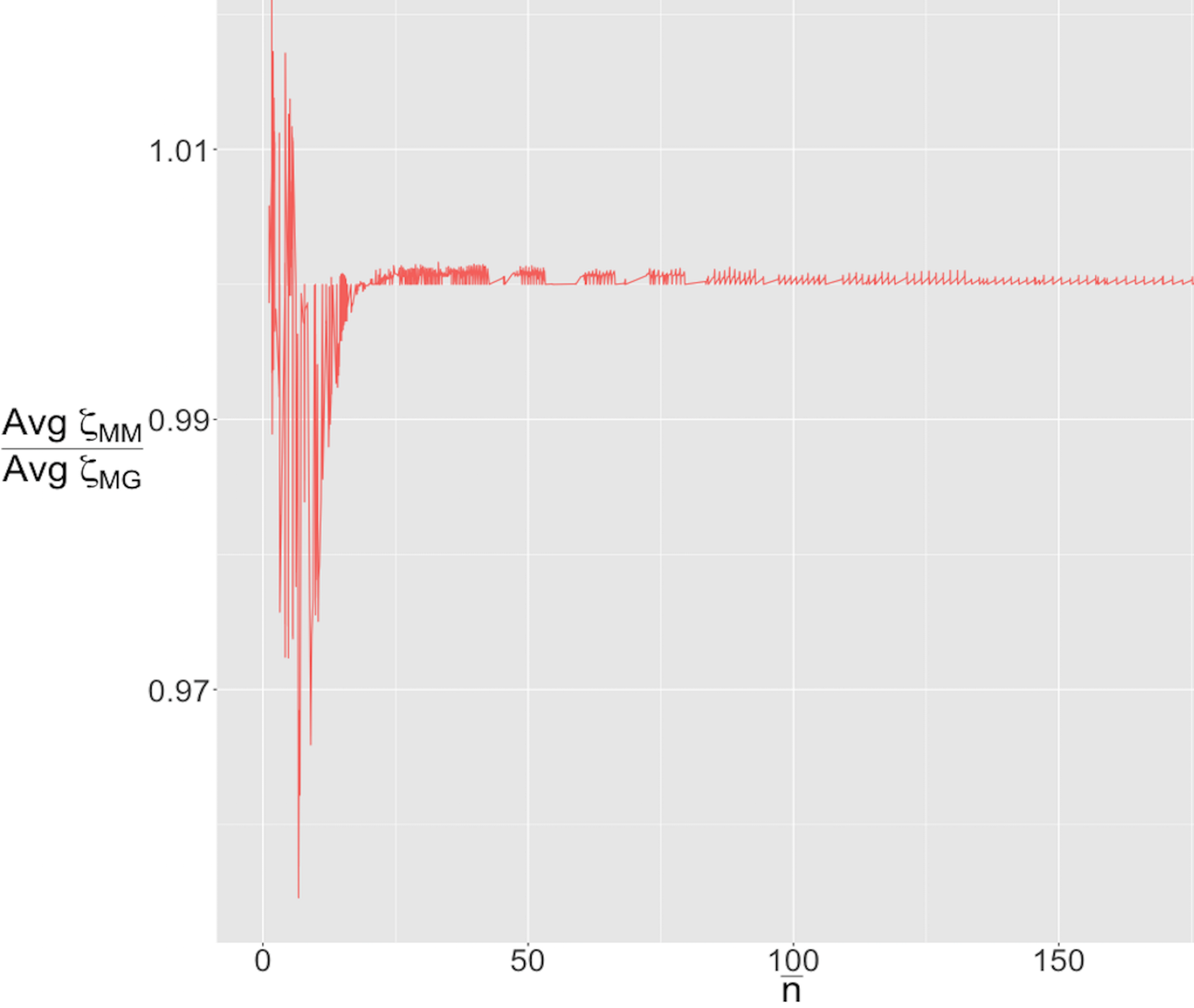}
  \caption{$\bar{N}=5$}
\end{subfigure}
\hfill
\begin{subfigure}{.46\textwidth}
  \centering
  \includegraphics[width=1.2\linewidth]{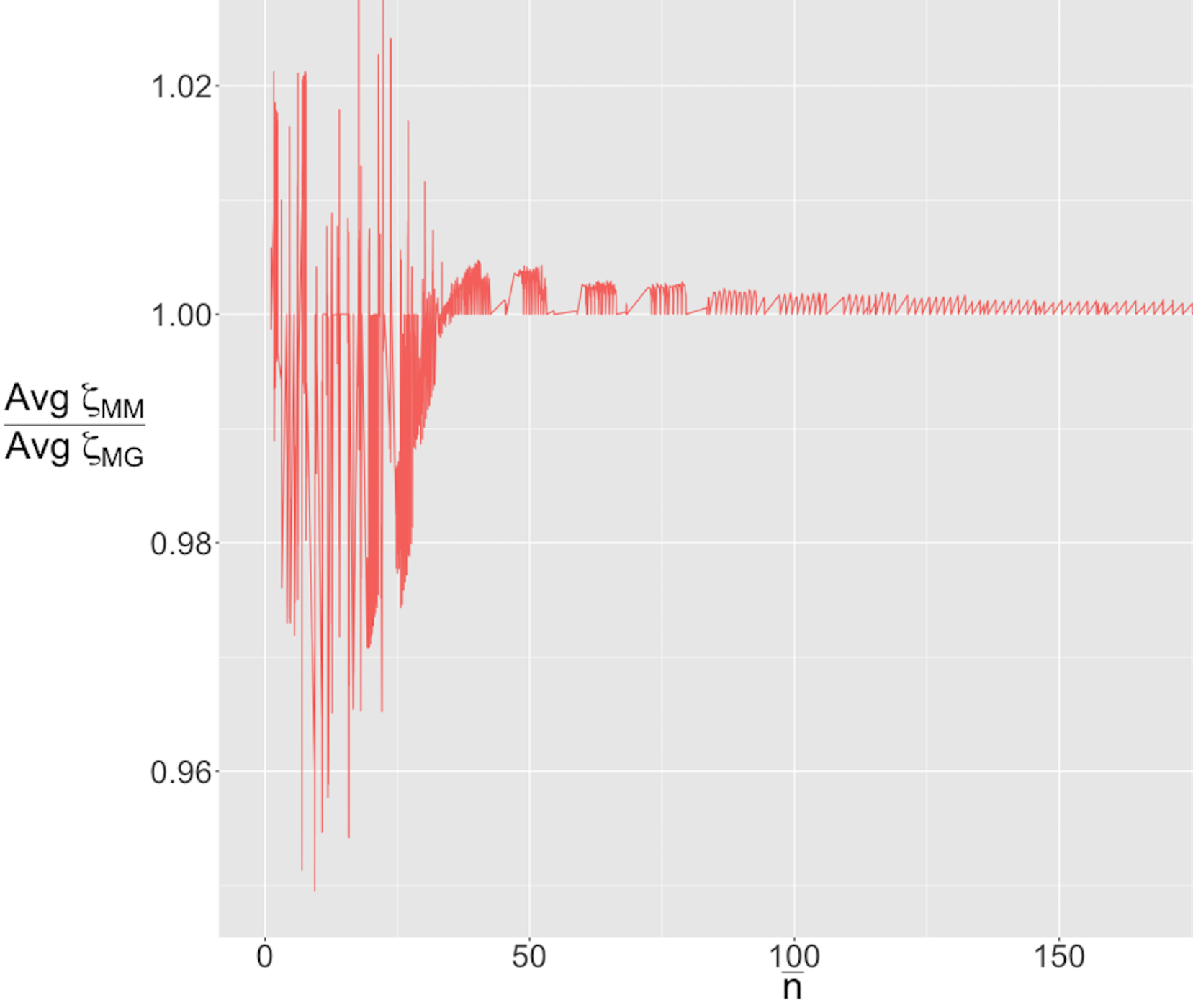}
  \caption{$\bar{N}=20$}
\end{subfigure}
\caption{Average $\zeta_{MM}$ over average $\zeta_{MG}$ vs. $\bar{n}$}
\label{fig:MMvsMG}
\end{figure}

\paragraph {Simulation.} We set $T$ to 200 and define ranges [0.7,15], [1,30], [1,23] and [5,20] for $\lambda$, $w_{max}$, $d_{max}$ and $\bar{N}$, respectively. We compare $\zeta_{MM}$ under different values of $\bar{N}$. We also average $\zeta_{MM}$, use the same simulations to run MG and average $\zeta_{MG}$. We take the ratio of both averages and plot it against $\bar{n}$ (Figure~\ref{fig:MMvsMG}). Preliminary results show that that for small $\bar{n}$, the algorithm, at higher $\bar{N}$, does slightly worse than at lower $\bar{N}$. However, the opposite is true for large $\bar{n}$. 

\paragraph{Future work.}  Further ideas are needed
to set the optimal choice for $\bar{N}$. We may want to look at the
percentage of times the algorithm chose to run MG over MLP in order to
monitor time complexity. Another idea would be to take hard instances
of MG into consideration and explore how to derive conditions such
that the algorithm switches to MLP in such cases.

\subsection{The Learning Modified Greedy Algorithm (LMG)}
\paragraph{Algorithm.} MG and MLP are memoryless~\cite{jez12},
i.e. the assignment of packets at time $t$ uses no information about
assignments at times $t' < t$.  
We introduce a non-memoryless modification to MG.
Recall that MG
compares $w_e$ to $\frac{w_h}{\phi}$. In the
learning MG (LMG) algorithm, we try to make use of the past
performance of MG, in order to replace $\phi$ by a more suitable
divisor. If $f$ defines the frequency of learning, then every $f$ steps, we calculate the throughput at the current divisor, $\phi^*$. Then we search for a divisor, $\phi_{better}$, which lies in the vicinity of $\phi^*$, but yields higher throughput using the previous data. Finally, we proceed with the new divisor $\phi^{*}_{new}$, given by ($\alpha \phi^{*} + (1-\alpha) \phi_{better}$)  for some smoothing factor $\alpha \in [0,1]$. The detailed procedure for LMG is given below: 

\paragraph{Step 0.} Set the frequency of learning, $f$, i.e. the time window needed to define a learning epoch. Then run MG and use the following procedure every $f$ steps in order to replace the divisor, $\phi$, by a sequence of better divisors as follows:

\begin {enumerate}
\item Generate a sequence of divisors $\phi_i$'s starting at the current divisor $\phi^{*}$ and having jumps of $\pm 0.05$, without going below 1 or above 2.5. For instance, if $\phi^{*}$=1.62, we generate the following sequence: 1.02, 1.07, $\hdots$, 1.57, 1.62, 1.67, $\hdots$, 2.47.

\item Start with the throughput associated with $\phi^{*}$ and move left in our generated sequence. At each $\phi_i$, we calculate the throughput of MG on the previous data. We keep moving to subsequent divisors as long as there is an increase in throughput. Next, we do the same with divisors to the right. Given a left endpoint and a right one, we choose the divisor associated with the higher throughput and denote it by $\phi_{better}$. Some toy examples are shown in Table~\ref{tab:LMG}. For simplicity, we only observe the weighted throughput for four values of $\phi$.

\begin{table}[ht]
\centering
 \begin{tabular}{c || c c c c || c} 
 \hline
 Thruput/$\phi_i$  & 1.57 & $\phi^{*}$=1.62 & 1.67 & 1.72 &   $\phi_{better}$\\  
 \hline\hline
Case 1  & 2& 4& 4& 2&  1.62 \\ 
 \hline
Case 2  & 2& 4& 2& 6&  1.62 \\ 
 \hline 
 Case 3  & 6& 4& 5& 7 & 1.72 \\ 
 \hline
 Case 4  & 4& 4& 4& 4 & 1.62 \\ 
 \hline
 Case 5  & 6& 4& 3& 7 & 1.57 \\ 
 \hline

\end{tabular}
 \caption{Examples for choosing $\phi_{better}$}
 \label{tab:LMG}
\end{table}

\item The new divisor $\phi^{*}_{new}$ is given by smoothing $\phi^{*}$ with $\phi_{better}$, i.e. for some $\alpha \in [0,1]$ $$\phi^{*}_{new} = \alpha \phi^{*} + (1-\alpha) \phi_{better}$$
\end{enumerate}

\paragraph {Simulation.} We use the same parameter space as
in~\ref{subsec:MM}. We set $f=\max(0.1*T,
\frac{30}{\min(1, \lambda)})$ and for simplicity, $\alpha= 0.5$. The choice for $\alpha$ in general must ensure that the process of finding an
optimal divisor $\phi$ does not generate a jumpy sequence of
divisors. Our analysis for 8000 sampled scenarios shows that LMG outperforms MG
83.3\% of the time. The range of the improvement is [-0.6\%, 2.8\%],
implying that LMG brings as much as 2.8\% increase in the ratio over
MG. Performance is worse when the sequence of $\phi^{*}_{new}$'s is
around 1.618, implying that LMG is picking up on some noise and
should not change the divisor.  

\paragraph{Future work.} One can avoid this noise by statistically testing and justifying the significance of changing the divisor. In terms of time complexity, LMG is not slower than MG, as it can be done while the regular process is running. Finally, no direct conclusion is made about a threshold on the number of packets beyond which LMG is particularly effective. Further analysis could yield such conclusion, thereby indicating at which instances LMG should be used. 

\subsection {The Second Max Algorithm (SMMG)}
\paragraph{Algorithm.} Inspired by the dummy packet (DP) algorithm discussed in~\cite{li07} for cases of non-agreeable deadlines, we realize the importance of extending the comparison to a pool of more than two packets. The
key idea in SMMG is to prevent the influence a single heavy packet may have on
subsequent steps of the algorithm. We try to find an early-deadline
packet that is sufficiently large compared to the heaviest packet. We
set a value for $p \in (0,1)$ and the iterations are as follows: 


\begin {enumerate}
\item If MG chooses $e$, send $e$ and STOP.
\item Else find the earliest second largest packet in the buffer, denoted by $s$. 
\item If $d_s < d_h$ and $ w_s \geq \max(w_e, p*w_h)$, send $s$. Else send $h$.
\end {enumerate}

The intuition here is that sending packet $e$ is always a good choice, so we need no modification. However, we limit over-choosing packet $h$ by finding the earliest second-largest packet $s$. The concern is that keeping $s$ in the buffer may bias the choice of the packets. Hence, we send $s$, if its weight is significant enough, in order to eliminate its influence and keep the possibility of sending $h$ for a subsequent iteration (as $h$ expires after $s$). To evaluate that $w_s$ is significant enough, we verify that it exceeds $w_e$ (otherwise, we should have sent e), as well as $p*w_h$, a proportion of $w_h$. Note that for instance, if $p=0.95$, it means that SMMG is very conservative allowing the fewest modifications to MG.  

\begin{figure}[ht]\centering 
\includegraphics[width=0.55\linewidth]{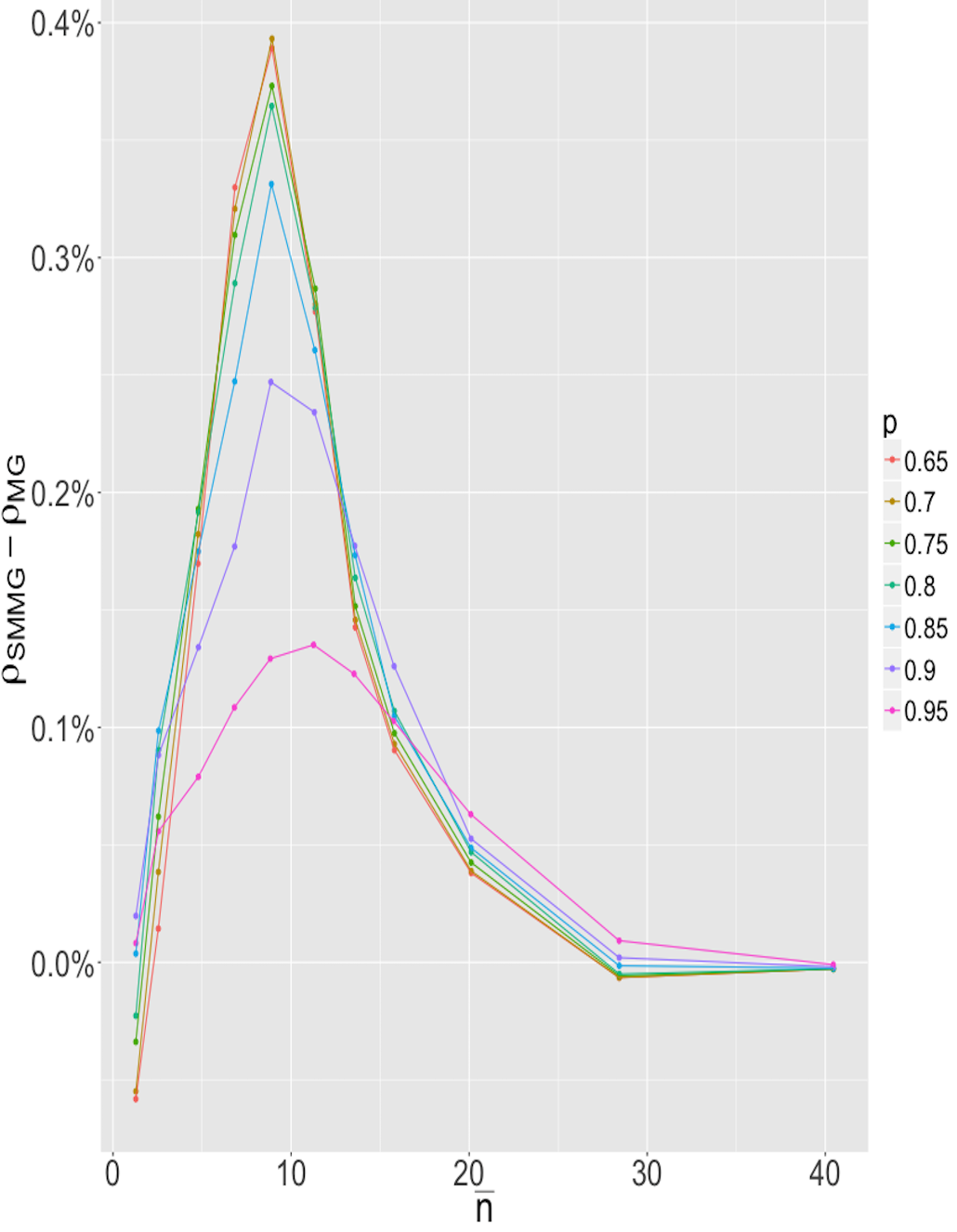}
\caption{($\rho_{MG}$ - $\rho_{SMMG}$) vs. $\bar{n}$, colored by p}
\label{fig:SMMG}
\end{figure}

\paragraph{Simulation.} We use the same parameter space as in~\ref{subsec:MM} and try values for $p$ as follows: 0.65, 0.75, 0.85, 0.95. Figure~\ref{fig:SMMG} plots the improvement of SMMG over MG ($\rho_{MG}$ - $\rho_{SMMG}$) vs. $\bar{n}$, colored by $p$.  As expected, the lower the value of $p$, the bigger the deviation from MG. At very low $\bar{n}$, we see that applying SMMG is not useful, however, as $\bar{n}$ increases, the improvement remains positive. At all values of $p$, the improvement is at its highest when $\bar{n}$ is between 8 and 12 packets. Hence, SMMG is useful when $\bar{n}$ is in the vicinity of the interval between 4 and 17. Whether this is a significant result, depends on the nature of our problem. Even if $p=0.95$, the minimum improvement within that interval is around 0.8\%. However, the maximum improvement is 1.5\% (at $p=0.95$). 

\vspace{-24pt}
\section {Model Discussion}
\label{sec:MD}

The two-node model with no buffer limitations clearly does not capture all aspects of 
realistic network models.  Thus, in this section, we will consider a
multi-node model and also the case of finite buffer capacity.

\paragraph{Multi-node model.} In the previous analysis, we only
considered a two-node-system, namely the source and the target
nodes.  In order to understand multinode systems, we consider first 
a three node system, that is a
system of two tandem  queues and see how the throughput behaves.

Assume the arrival rate at node 1 is $\lambda_1$ and that each packet
has to be sent to visit node 2 then reach node 3 by its deadline. Some
packets will be lost at node 1, as they expire before being sent. Node
2 will, hence, have less traffic. We assume as before that we can send
at most one packet per node at each time step. Within this framework,
we are interested to know whether this setup results in a
deterministic online algorithm of better performance. Our simulation
shows that node 2 either has the same throughput as node 1 or
lower. After tracing packets, it turns out that this is a fairly
logical result because node 2 only receives packets from node 1. The
packet will either expire at stage 2 or go through to node 3. So the
throughput can only be at most the same as that of node 1.

The following minor adjustment slightly improves the performance at
node 2: Each arriving packet has a deadline to reach node 3, denoted
by $d$. We introduce a temporary deadline for that packet to reach
node 2 by $d-1$. This modification guarantees that we only send packets to
node 2 if after arriving at node 2 there is at least one more time
unit left to its expiration in order to give the packet a chance to
reach node 3. Here is a trivial example: A packet arrives with
deadline 7, i.e. it should arrive at node 3 by 7. Before the
adjustment it was possible for this packet at time 7 to be still at
node 1 and move to node 2, then be expired at node 2 and get
lost. After the adjustment, this packet will have a deadline of 6 for
node 2. So if by time 6, the packet hasn't been sent yet, it gets
deleted from the buffer of node 1 (one time step before its actual
deadline). This adjustment improved the throughput of node 2 to be
almost equal to that of node 1 because the arrival rate at node 2 is
at most 1 (at most one packet is sent from node 1 at each time
step). So node 2 is usually making a trivial decision of sending the
only packet it has in its buffer.

In conclusion, our model
implicitly imposes a restriction on the maximum possible throughput at internal
nodes, hence, making the multi-node model, where only one packet is
sent at each time step, an uninteresting problem. In~\ref{sec:Conc}, we give, however, a few future directions for more interesting extensions.

\begin{wrapfigure}[10]{r}{0.45\textwidth}
\vspace{-20pt}
\includegraphics[width=\linewidth]{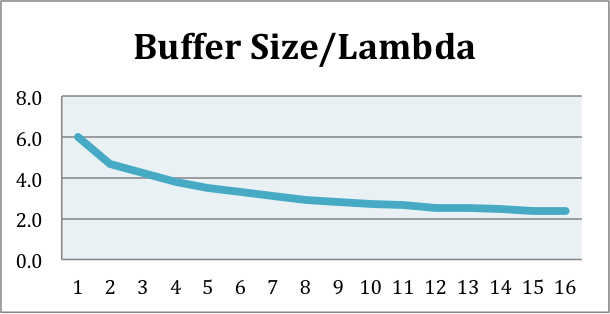}
\caption{ $b/\lambda$ vs. $\lambda$}
\label{fig:finite}
\end{wrapfigure}

\paragraph{Finiteness of Buffer.} Throughout this paper, we chose not to put any restrictions on the buffer capacity and therefore we now look to verify whether the finiteness of the buffer has a major effect on the algorithm performance and its bounds. We run a set of experiments where, given $\lambda \in [2, 100]$, we find the corresponding buffer size $b$, such that the probability of exceeding $b$ is one in a million. We create an index plot for the ratio of $b/ \lambda$ (Figure~\ref{fig:finite}) and conclude that imposing a buffer size is unnecessary. Even when $\lambda = 100$, the buffer size needs to be about 1.53 as much, i.e. 153.  Furthermore, for the interesting values of $\lambda$ used throughout this report, a buffer size of around 30 would be more than sufficient. We conclude that imposing a capacity limit on the buffer is not necessary as choosing a
reasonable buffer size should not affect the algorithm's
performance.

\section {Conclusion}
\label{sec:Conc}
In this paper, we consider several old and new packet scheduling algorithms.
By 
analyzing the empirical behavior of MG, we observe that MG
chooses packet $h$ over $e$ too frequently. We
therefore develop a new algorithm, MLP, which
mimics the offline algorithm and gives higher attention to early
deadline packets.  We then show that on a wide variety of data, including
uniform and bimodal distributions, 
MLP is slower, but has a better empirical competitive ratio than MG. (This is in contrast to
the worst-case analysis where MG has a better competitive ratio.)

We then propose three new algorithms that may offer an improvement in
empirical performance, as they combine features of both
algorithms. MM, at each time step, chooses between using MG or MLP in
order to make a decision on the packet to send. LMG learns from
previous behavior to correct the divisor used in MG, while SMMG is
motivated by the idea of influential packets in extending the
comparison to a pool of three packets, namely $e$, $h$ and $s$. The
improvements for these algorithms are small, yet encouraging for
further analysis. Moreover, it is important to consider extensions for
the network model and run the algorithms on one where induced
correlations are captured by more realistic distributions that are not
i.i.d.  Contrasting the behavior of any of the algorithms mentioned in
this paper on an actual router, rather than a simulated environment,
would also be important to consider.

Several interesting future directions remain.  One important extension would
be a multi-node model.  We showed how the straightforward extension does not
yield much insight but other extensions may be more interesting.
For example, one could have nodes that process at different rates; this would
prevent the first node from being an obvious bottleneck.
Another possibility is to allow feedback, that is, if a packet expires somewhere in the multi-node system, it could return to the source 
to be resent.
A final  possibility is for each packet to have a vector of deadlines, one per node, so
that different nodes could be the bottleneck at different times.

\section*{Acknowledgments}

The authors would like to thank Dr. Shokri Z. Selim and Javid Ali.


\mbox{ }
\newpage
\mbox{ } 
\appendix
\section{Appendix}
\label{sec:app}
\counterwithin{figure}{subsection}
\setcounter{table}{0}
\renewcommand{\thetable}{A\arabic{table}}

\subsection{General Procedure for Simulations in~\ref{sec:CA}}
\label{app:CA}
Our simulations try to avoid any initial bias by starting with a nonempty system. This is the role of the variable $\kappa$ which "warms up" the system; that is the simulation will take place for $\kappa$ time steps before collecting any data. This warmup period allows the content of the buffer to adjust to normal running conditions for the system. The general simulation procedure is as follows:

\begin {enumerate}
\item Identify a parameter range for the parameters $T$, $\lambda$, $w_{max}$ and $d_{max}$.
\item Randomly select a parameter combination within the parameter space. 
\item Set $\kappa=\max(40, 0.4T)$. 
\item Generate inputs for $T+\kappa$ time steps using the chosen parameter combination.
\item Find the optimal offline solution, for the last T steps only, i.e. sum of weights of packets sent from $t=\kappa+1$ to $t=\kappa+T$ and let that be our $\zeta_{OFF}$.

\item Run MG and record the weighted throughput for the last T steps only.
\item Repeat Step 6 five times and let our $\zeta_{MG}$ be the average weighted throughput over the five runs.

\item Run MLP and record the weighted throughput for the last T steps only.
\item Repeat Step 8 five times and let our $\zeta_{MLP}$ be the average weighted throughput over the five runs.

\item Compute $\rho_{MG}$, $\rho_{MLP}$, as well as $\hat{\rho}=\frac{\zeta_{MLP}}{ \zeta_{MG}}$.
\item Repeat Steps 2-10 500 times.
\end{enumerate}

Note that the input generation in Step 3 has been described in~\ref{sec:Introduction}.

\subsection{Predictive Model}
\label{app:predM}
We construct a gradient boosted tree (GBT) predictive model on the results for inference purposes. The predictors were $d_{max}$, $w_{max}$, $T$, $\lambda$ and whether MG or MLP were used and the response was $\rho$ of the respective algorithm. In terms of variable importance (as measured by the model and stated as a percentage), $\lambda$ and $T$ turn out to be the two most important parameters.

\vspace{-13pt}
\begin{table}
\centering
 \begin{tabular}{|c || c c c c | c c|} 
 \hline
 \textbf{Feature} & $\mathbf{\lambda}$ & $\mathbf{T}$ &  $\mathbf{d_{max}}$ & $\mathbf{w_{max}}$ & \textbf{MLP?} & \textbf{MG?} \\
 \hline
 \textbf{Variable Importance}& 41\%  & 15\% & 8\%& 3\% & 17\% & 16\%\\  
  \hline

\end{tabular}
 \caption{Variable Importance for a GBT predictive model}
\end{table}

\begin{figure*}\centering

\begin{subfigure}[b]{.45\textwidth}
  \includegraphics[width=1.25\textwidth]{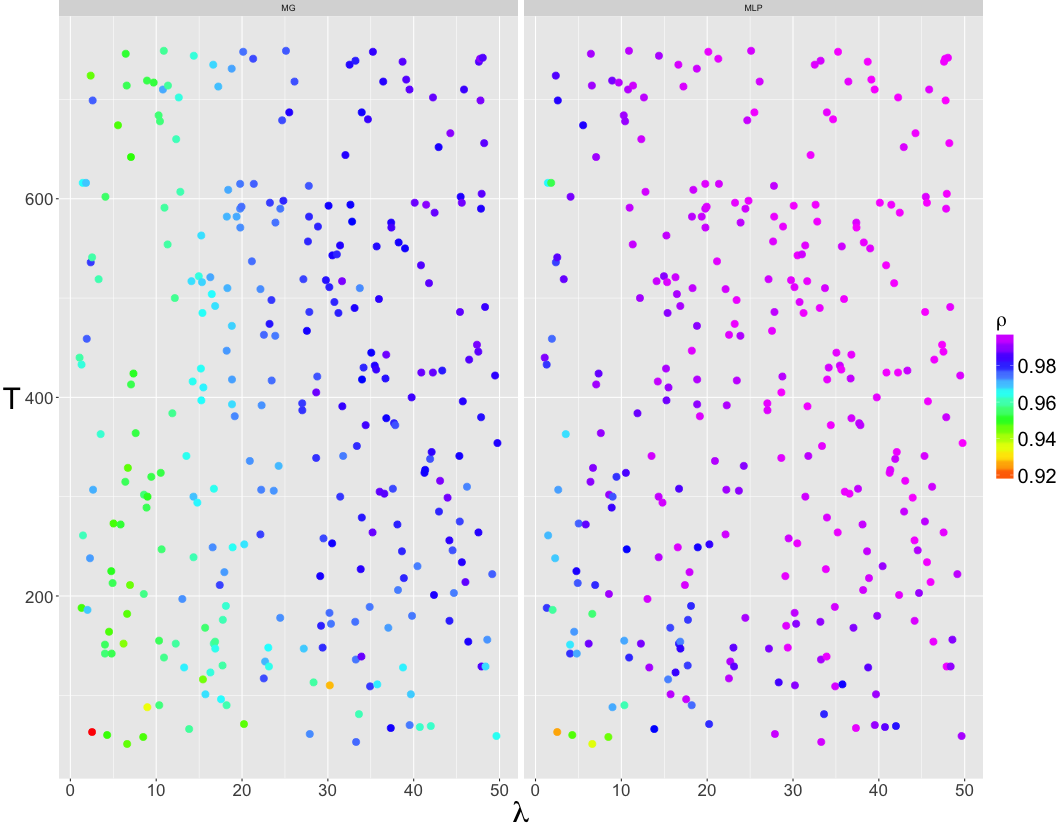}
  \caption{Colored by $\rho_{MG}$(left) and $\rho_{MLP}$(right)}
  \label{fig:Comp2TL}
\end{subfigure}
\hfill
\begin{subfigure}[b]{.45\textwidth}
  \includegraphics[width=1.25\textwidth]{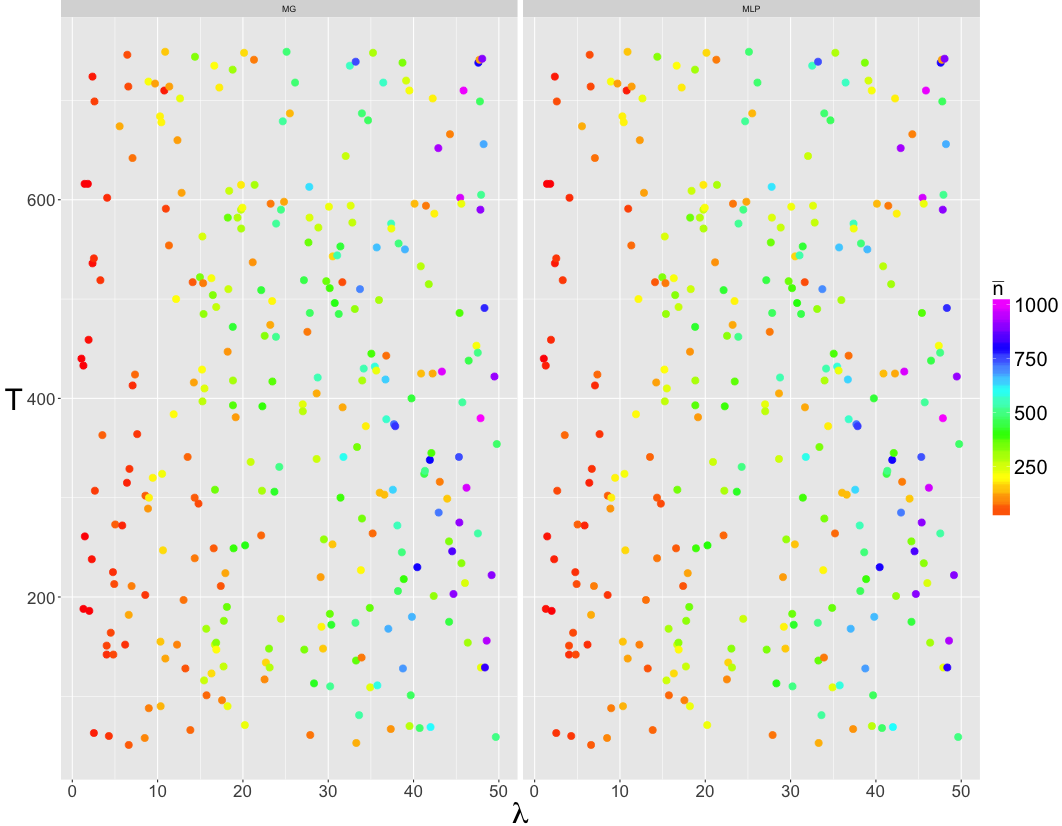}
  \caption{Colored by $\bar{n}_{MG}$(left) and $\bar{n}_{MLP}$(right)}
  \label{fig:Comp3TL}
\end{subfigure}
\caption{$T$ vs. $\lambda$}
\label{fig:Comp23}
\end{figure*}

\vspace{-26pt}
Figure~\ref{fig:Comp2TL} plots both model parameters for each of the algorithms, and colors them by $\rho$. This allows us to view the behavior of the ratio in a multivariate manner: We see here that there is a point where the ratios tend to level off, approximately (as can be seen by the constancy of the colors) around $\lambda>20$ and $T>200$. We looked to see if this was perhaps a result of the average number of packets in the buffer but saw slight correlation, as is evident in Figure~\ref{fig:Comp3TL}.

\begin{figure}\centering

\begin{subfigure}{.45\textwidth}
  \centering
  \includegraphics[width=\linewidth]{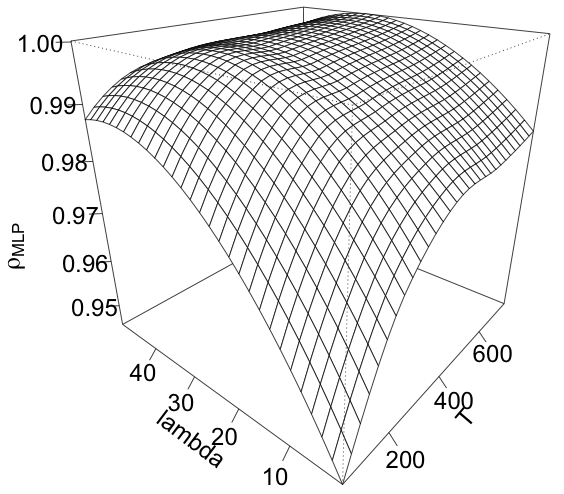}
  \caption{acutal $\rho_{MLP}$}
  \label{fig:Ratio3DMLP}
\end{subfigure}
\hfill
\begin{subfigure}{.45\textwidth}
  \centering
  \includegraphics[width=\linewidth]{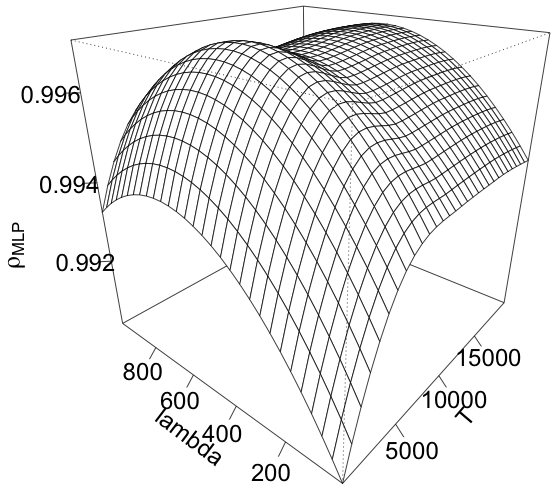}
  \caption{predicted $\rho_{MLP}$}
  \label{fig:PredRatio3DMLP}
\end{subfigure}

\begin{subfigure}{.45\textwidth}
  \centering
  \includegraphics[width=\linewidth]{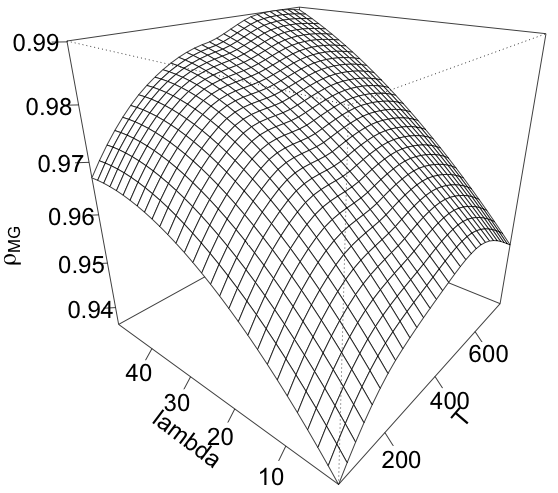}
  \caption{acutal $\rho_{MG}$ }
  \label{fig:Ratio3DMG}
\end{subfigure}
\hfill
\begin{subfigure}{.45\textwidth}
  \centering
  \includegraphics[width=\linewidth]{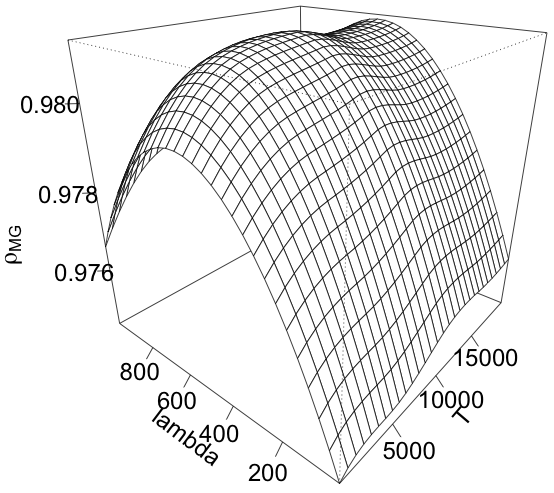}
  \caption{predicted $\rho_{MG}$ }
  \label{fig:PredRatio3DMG}
\end{subfigure}

\caption{3D locally smoothed plot of $\rho$ vs. $T$ and $\lambda$}
\label{fig:3DMLP}
\end{figure}

Next, we perform predictions from the model for a much larger range of $T$ and $\lambda$ (risking extrapolation issues). In Figure~\ref{fig:PredRatio3DMLP}, $\rho_{MLP}$ is well behaved for large values of $T$ and $\lambda$: there seems to be a dip but that may be due to randomness. Furthermore, the ratios do not dip as sharply as for MG in Figure~\ref{fig:PredRatio3DMG}. This leads us to investigate more the behavior of $\rho_{MG}$ for larger values of $\lambda$, which leads us to Scenario 4 where values for $\lambda$ are extended to $(0.5,250)$ and the same analysis is run for MG.

The results for the behavior of $\rho_{MG}$ vs. $\lambda$ shows an increasing performance under Scenario 4. We are 99\% confident that $\zeta_{MG}$ is at most 98.69\% of $\zeta_{OFF}$. Plots against other parameters show that the performance of MG, all else constant, would get better with larger $T$ or lower $d_{max}$ and only slightly better for larger $w_{max}$. A gradient boosted tree predictive model again shows that $\lambda$ and $T$ are the two most important variables. As $\lambda$ and $T$ get larger, $\rho_{MG}$ gets better.

\section{Figures}
\label{app:fig}

\subsection {Parameter Effect on MLP Behavior}
\label{app:MLP}
\vspace{-18pt}
\begin{figure*}[h!]\centering

\begin{subfigure}{.45\textwidth}
  \centering
  \includegraphics[width=1.1\linewidth]{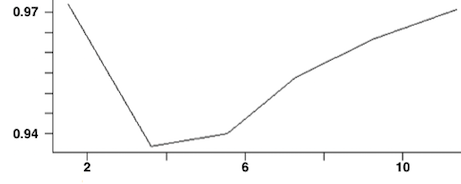}
\caption{ $\lambda$ focuses on sensitive values}
\label{fig:MLP2}
\end{subfigure}
\hfill
\begin{subfigure}{.45\textwidth}
  \centering
  \includegraphics[width=1.1\linewidth]{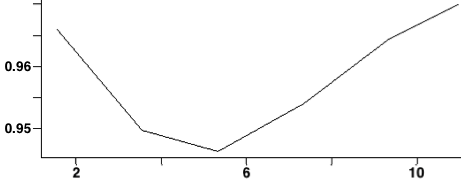}
\caption{$w_{max}$ reflects narrow weight ranges\\}
\label{fig:MLP3}
\end{subfigure}

\begin{subfigure}{.45\textwidth}
  \centering
  \includegraphics[width=1.1\linewidth]{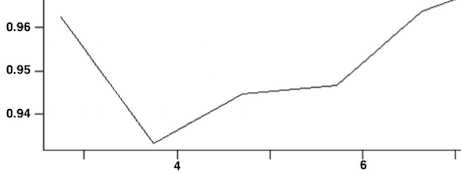}
  \caption{longer $T$}
  \label{fig:MLP4}
\end{subfigure}
\hfill
\begin{subfigure}{.45\textwidth}
  \centering
  \includegraphics[width=1.1\linewidth]{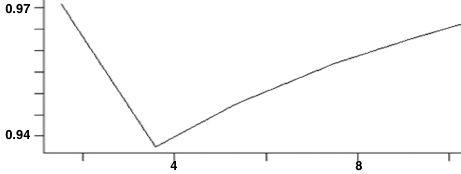}
  \caption{smaller $d_{max}$}
  \label{fig:MLP5}
\end{subfigure}

\caption{$\rho_{MLP}$ vs. $\bar{n}$}
\label{fig:MLP2345}
\end{figure*}

\mbox{ } 

\newpage

\subsection{$\rho_{MLP}$(red) and $\rho_{MG}$(green) vs. different parameters} 
\label{subsec:Ratio}
\mbox{ }

\begin{figure*}[h]
\centering
 \begin{subfigure}{.45\textwidth}
  \centering
  \includegraphics[width=1.2\linewidth]{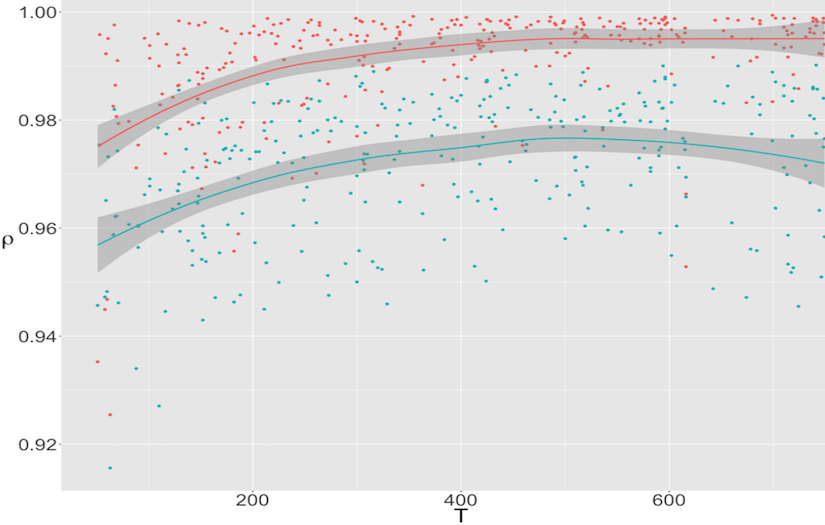}
  \caption{$T$}
  \label{fig:CompT}
\end{subfigure}
\hfill
 \begin{subfigure}{.45\textwidth}
  \centering
  \includegraphics[width=1.2\linewidth]{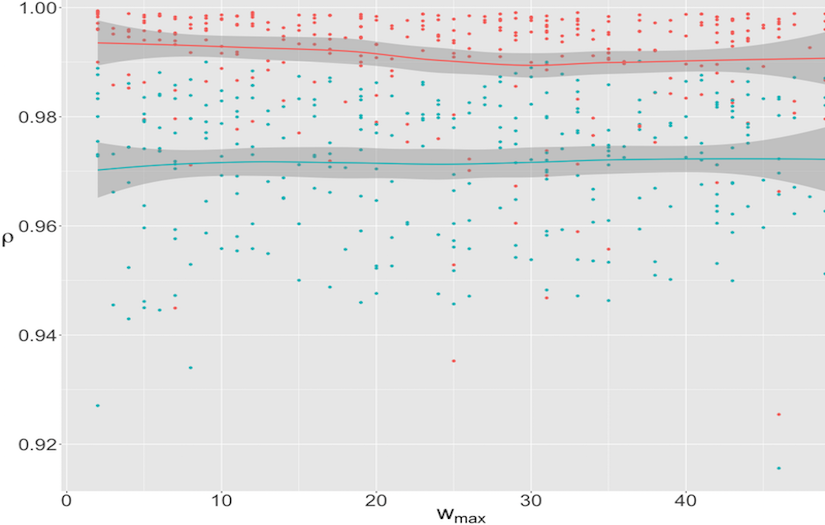}
  \caption{$w_{max}$}
  \label{fig:CompW}
\end{subfigure}

 \begin{subfigure}{.45\textwidth}
  \centering
  \includegraphics[width=1.2\linewidth]{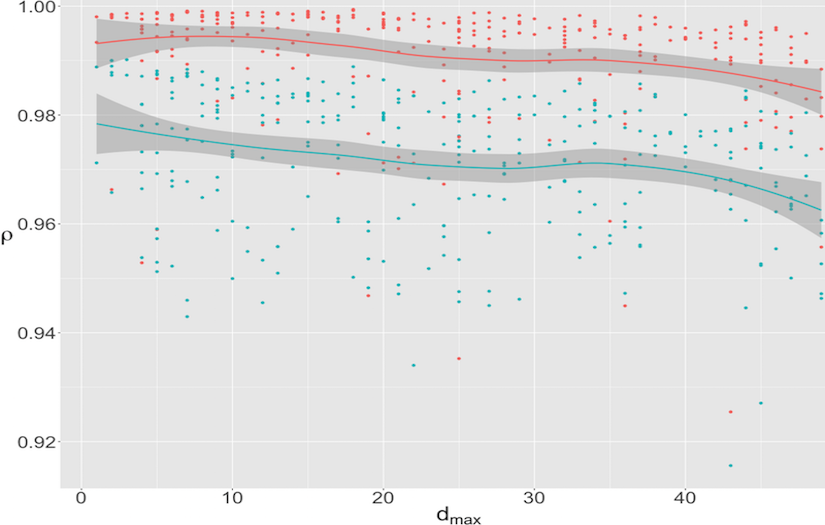}
  \caption{$d_{max}$}
  \label{fig:CompD}
\end{subfigure}
\hfill
 \begin{subfigure}{.45\textwidth}
  \centering
  \includegraphics[width=1.2\linewidth]{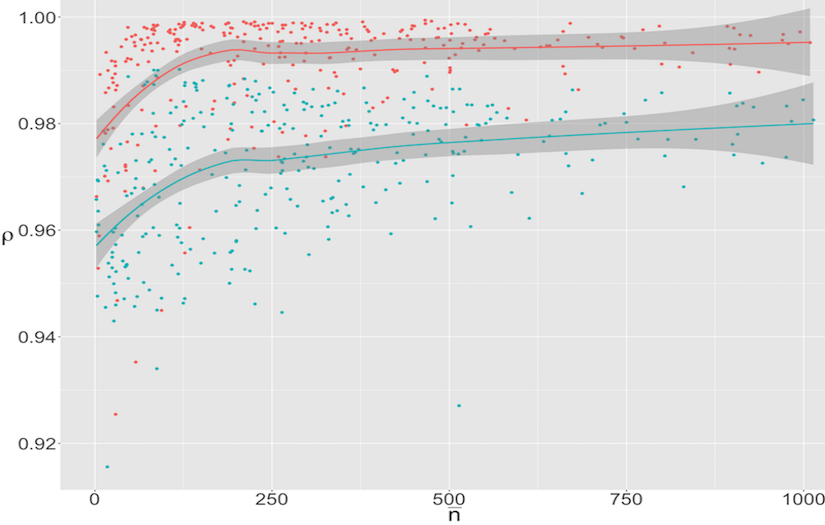}
  \caption{$\bar{n}$}
  \label{fig:CompN}
\end{subfigure}

\caption{$\rho_{MLP}$(red) and $\rho_{MG}$(green) vs. different parameters}
\label{fig:COMP}
\end{figure*}

\mbox{}

\end{document}